# AERODYNAMIC STUDY OF LEADING-EDGE PROTUBERANCE TO ENHANCE THE PERFORMANCE OF NACA 0009 BLADE

Chaitanya Kumar Konda[a*], Vidyashankar. S[b], Ulavish. V. S[b], Sachin. A. M[b], Mahesh. K. Varpe[c]

[a]Department of Mechanical Engineering, Indian Institute of Technology, Ropar, Rupnagar-140001, Punjab, India
[b]Independent Researchers
[c]Department of Aerospace and Automotive Engineering, M S Ramaiah University of Applied Sciences, Bangalore-560058, Karnataka, India

## ABSTRACT

*Symmetric NACA airfoils tend to undergo abrupt stall characteristics at higher angle of attacks. The abrupt stall has deteriorating effect on lift as well as the efficiency of the airfoils. Abruptness in stall restricts the airfoil to operate only at lower angle of attacks. So, in order to improve the efficiency of airfoils at higher angle of attacks and make it suitable for operation over higher range of angle of attacks, there are many flow control techniques. One such technique is addition of leading-edge protuberance. Leading-edge protuberances are the leading-edge modification of the wing. Leading-edge of the wing is modified with sinusoidal structural modification. This modification has two parameters i.e., Pitch and Amplitude. Many configurations of the protuberances can be obtained by changing the Pitch to Amplitude ratio of the protuberance. In the present work, the Reynolds number is 50k for NACA 0009. The Pitch to Amplitude ratio is varied from PAR1 to PAR27. PAR6 is found to be the better case which has higher lift and efficiency in the post-stall angle of attacks. At the deep stalling AOA of the baseline, i.e., at 13.6°, PAR6 is found to have the highest increase in lift and efficiency compared to the other post stalling AOAs with it having around 39.6% more lift and 27.3% more efficiency compared to the baseline. At the higher angle of attacks, normal airfoils undergo flow separation over its entire suction surface, while the modified (PAR6) case delays the flow separation. The main mechanism behind delaying the flow separation is due to counter-rotating vortices formed due to leading edge modification. The counter-rotating vortices help in intermixing of the flow and mix the low momentum flow with high momentum flow. This helps in delaying the flow separation over the suction surface over the modified blade, which in turn helps in retaining the lift as well as the efficiency of the airfoil at higher AOAs.*

## NOMENCLATURE

| | |
|---|---|
| AOA | Angle of Attack, degrees |
| BLT | Boundary layer thickness, m |
| CFD | Computational Fluid Dynamics |
| $C_d$ | Drag Coefficient |
| $C_l$ | Lift Coefficient |
| $C_l/C_d$ | Efficiency |
| Cp | Pressure Coefficient |
| Cpo | Total Pressure Coefficient |
| Cps | Static Pressure Coefficient |
| PAR | Pitch-to-Amplitude Ratio |
| $PL_\alpha$ | Peak Lift AOA, degrees |
| PS | Pressure Surface |
| $R_N$ | Reynolds Number |
| SS | Suction Surface |
| SST | Shear Stress Transport |
| SIMPLEC | Semi-Implicit Method for Pressure Linked Equations-Consistent |
| U | Free-Stream Velocity, m/s |

## 1. INTRODUCTION

The aerodynamics within the boundary layer plays a pivotal role in the performance of airfoils. It is very significant and it has to be taken great care of during the stall of an airfoil. The separation of boundary layer over an airfoil is a much-sophisticated area of investigation to comprehend the flow physics and the enhancement of aerodynamic performance. So, intensive experimental studies and sceptical techniques to harness the flow within the boundary layer have been done since decades. The advancement of computational fluid dynamics (CFD) with improved processing capacity of computers has accelerated the investigations in the boundary layer flow studies. Active flow and Passive flow control techniques serves as better methods to control the flow within the boundary layer region to enhance the performance of airfoil. As passive flow control techniques are easier to design and investigate; numerous methods have been proposed. Fixing the grit on both suction and pressure surface on airfoil, addition of sound in wind tunnel tests, increasing the surface roughness have been studied by Mueller et al [2]. Dimples on the suction surface, Vortex generators, riblets, protuberances on the leading edge of airfoil are the some of the passive flow control techniques. Introducing such techniques have influenced the formation of laminar separation bubble giving sudden rise in lift coefficient with a compromise in the maximum lift compared to the baseline profile of NACA $66_3$-018 [2]. Similar sudden rise in lift coefficients at higher angle of attacks and delay in the stall characteristics for the baseline profile of NACA 0020 with leading edge protuberances were observed by Miklosovic et al [3].

The basic idea of fitting the protuberance to the leading edge of the wing stems from the observation of pectoral flippers of humpback whale (*Megaptera novaeangliae*) [4]. Fish et al [4] studied an unusual pattern present on the pectoral flippers of humpback whale. They were among the first to correlate the hydrodynamic performance of the pectoral flippers in whale with unusual leading-edge patterns. These leading-edge patterns, now known as protuberance or tubercles act as passive flow control devices. Unlike other whales, these whales are very agile and are able to perform deep underwater maneuvers, despite of its large size. They concluded that the pattern of the leading-edge acts as passive flow control device, thereby producing better hydrodynamic performance at higher angle of attack. Since then, a series of research and experimental investigations have been done to explain the flow phenomenon of this unusual pattern [1,5,6,7] to extend its application to engineering interests.

[*]Corresponding author
Email address: chaitanyak2306@gmail.com (Chaitanya Kumar Konda)

Johari et al., Cai et al., [1, 8], conducted an experimental study on NACA $63_4$-021 airfoils which closely resembles the cross section of Humpback whale flippers. The study of wings with varying amplitude and wavelength of the leading-edge protuberance found that wing does not stall as rapidly as the baseline wing. It was also concluded that the presence of protuberance had more aerodynamic performance benefits in poststall region, although the drag was minimally affected by the leading-edge protuberance in the poststall region. Hansen et al., [5] concluded from the experimental investigations of the leading-edge protuberance on NACA 0021 and NACA 65-021 airfoils that the sinusoidal protuberance having an optimal wavelength had better performance compared to the baseline in post stall region. The insight of pitch to amplitude ratio (A/λ ratio) of protuberance was presented on a lighter note.

Authors of [1, 5, 9, 10] have mentioned that decreasing amplitude and wavelength of protuberance enhanced the performance of the airfoils. Custodio et al [6] conducted an experimental study on the performance of NACA $63_4$-021 airfoils at various Reynolds numbers. The amplitude was varied from 2.5 - 12% of chord while wavelength was varied from 25 - 50% of chord. Various combinations of wavelengths and amplitude with leading-edge protuberance were studied through water tunnel experiments. The amplitude of the protuberance was found to reduce the span efficiency of the wings. Skillen et al [7], performed a numerical simulation of infinite span wings of NACA 0021 profile using large eddy simulations. The modified wing had better gradual stall characteristics compared to the baseline wing. Zhang et al [11], conducted an experimental study on DU00-W2-401 airfoil. They concluded that leading-edge protuberances can effectively suppress flow separation for an airfoil in stall regions, moreover, the airfoil having smaller amplitude and wavelength perform significantly better in flow separation control at higher angle of attack.

Miklosovic et al [12] conducted experimental study on NACA 0020 airfoil and they modified by fitting the protuberance into full-span and semispan of the wing. By their experimental study for full span model, they concluded that in prestall region the lift is reduced by 38% and in post stall region the lift is increased by 48% and the drag is reduced by 6% when compared to baseline. For semispan model in the prestall region the lift is reduced by 13% than the baseline. However, the maximum lift is increased by 4% compared to baseline. The similar observation was observed by Yoon et al [13] wings with waviness showed better performance in post stall zone. Protuberance works as the vortex generator in the leading edge of an airfoil. This helps the flow to attach over the wing surface, even though the smooth wing has lost flow attachment from the leading edge. S. Sudhakar et al [14] used S1223 airfoil to construct two modified wings with leading-edge tubercles, one with constant amplitude and wavelength throughout the span and the other with varying amplitude and wavelength. Experimental results at $R_N$ = 0.18 million showed improved post-stall aerodynamic performance with a delay in stall. The modified wing with varying amplitude and wavelength had a higher post-stall lift coefficient. Unlike baseline wing, the modified wings did not show any hysteresis loop. The study concluded that higher $R_N$ does not affect the stall angle but affects the lift and drag coefficients.

Huang et al [15] carried out an experimental study on airfoils with leading edge protuberance of various amplitude and wavelength which were based of SD8000 foils. They concluded that in the prestall region there is no much difference in the lift and drag characteristics but in post stall region modified airfoils showed better performance. Protuberance with combination of larger amplitude and smaller wavelength will leads to drop in lift coefficient due to drastic flow interference occurring at the leading-edge region. Cai et al [16] conducted numerical analysis on hydrofoils with leading edge protuberance having NACA $63_4$-021 as an underlying profile. The lift coefficient for smaller and larger amplitude foils is reduced by 10% and 20% respectively in the prestall region where as in post stall region lift is increased by 20% and 30% respectively compared to baseline. The flow visualization near wall showed that protuberance worked in pairs at high angle of attack, producing different forms of streamwise vortices and an attached flow over some valley section was noticed, leading to higher lift at post stall angles.

Bapat et al [17] computationally analyzed the performance of blade using protuberance of NACA $63_4$-021. In the prestall region, when compared to the baseline, the protuberance blade stalls earlier and the maximum lift coefficient is lower. In the post stall region, the performance of protuberance blade was increased by 34.7% compared to baseline. The drag was found to be less in poststall region because of the formation of vortices that helps in improving the momentum exchange in the boundary layer region, which significantly minimizes pressure drag. Malipeddi et al [18] conducted numerical analysis on airfoil with leading edge protuberance which is based on NACA 2412. From the results, they concluded that airfoil without modification has better lift in lower AOAs, but after crossing 12°, the lift for the modified airfoil has increased by 25%, and above 16°, the lift decreases gradually but is still 15% greater when compared to the baseline. The drag for the baseline was found to be quadratic with AOA, and showed rapid increase in drag after the stall angle. The modified wing had a drag which showed similar effect up to stalling angle, but after that they generated less drag and eventually became almost equal to the baseline. Vortices in the stream were created by protuberances along the leading edge. These vortices transported the higher momentum flow in the boundary layer, delaying the separation by keeping the flow attached to the wing's surface.

Sreejith et al [19] studied the effect of leading-edge protuberance on a thin E216 airfoil experimentally and numerically at a low Reynolds number. By the results obtained they commented that modified airfoils had an improved lift coefficient of 4.51% and a better gradual stall characteristic compared to the smooth airfoil. Van Nierop et al [20] used an elliptic wing with leading-edge bumps. The study was conducted numerically using Prandtl lifting line approach. They demonstrated how the boundary layer was delayed behind the bumps, which finally led to a gradual beginning of the stall. Asli et al [21], conducted a numerical study on the S809 airfoil by using SST $k$-$\omega$ turbulent model. They found that, based on the flow physics over the airfoils, leading-edge bumps act as vortex generators. Vortices have high levels of momentum thereby helping the flow to remain attached to the surface of the airfoil at higher AOAs.

Hendrickson et al [22] compared the aerodynamic coefficients of S1223 airfoil with protuberance with Selig et al [23] S1223 airfoil without protuberance and found that the lift coefficients for S1223 airfoil with protuberance in the pre stall and post stall regime were lesser than the S1223 airfoil without protuberance, and in the region close to the critical stall angle, the lift coefficient was found to be 50% lower than the flat airfoil. Though the drag coefficients of the serrated airfoil was found to be similar to the flat airfoil in the lower angle of attacks, as the angle of attack increased, the drag coefficients for the serrated airfoil continued to increase, and finally in the post stall region, it was found to be twice as large as that for the flat airfoil. The best and worst performing amplitudes for wavelengths of 0.5c and 0.5c were found to be 0.04c and 0.08c respectively in terms of better aerodynamic performance. The poor performance of the serrated airfoil is because of the early separation at the troughs and delayed separation at the peaks.

The study of protuberance and its effects on the stall has been an important topic of study since, many years. But the perfect characterization of protuberance for a class of airfoils has not been concluded yet. In this work, we do a parametric study of protuberance and its effects on the profile of NACA 0009 Blade.

## 2. METHOD AND NUMERICAL MODELLING

### 2.1 Geometry & Mesh

A C-shaped domain with hexahedral-structured meshes is created using ICEMCFD. The mesh has a 'C' domain, and the origin of the coordinate system is at the leading edge of the airfoil. The curved zone, as well as the zones directly above and below the air foil, are considered velocity inlets with free stream velocity, U = 0.81m/s, i.e., at $R_N$ = 50000. The coordinate systems' origin is at the leading edge of the blade. Flow inlet and outlet boundaries are located 10C upstream and 15C downstream, respectively. The top and bottom wall boundaries are located by 10C from the origin as shown in Figure 1. The material used for our numerical investigation for the entire domain is Air, with a density of 1.1649 kg/m$^3$ and viscosity of 1.858 * 10$^{-5}$. Wall Y+ is found to be nearly equal to 1.

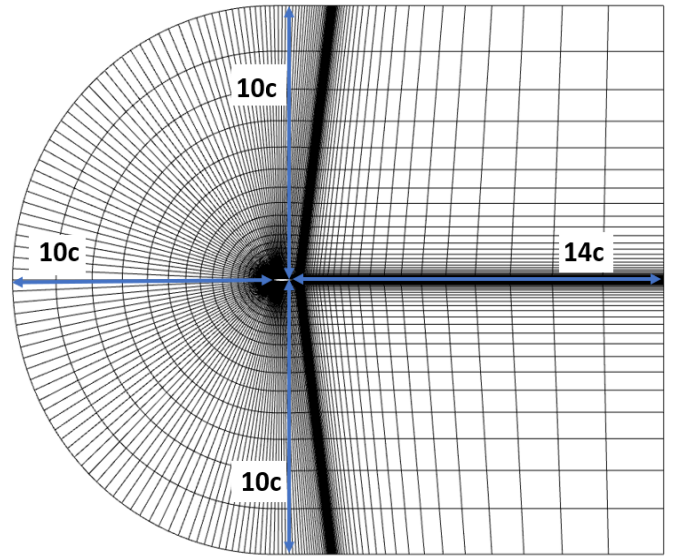

**FIGURE 1:** Computational Domain with Mesh Numerical

analysis is done using Ansys Fluent. The κ-ω SST turbulence model is used. The boundary condition prescribed at inlet is velocity inlet with periodic condition for the side walls, while PS and SS are regarded as walls. The downstream boundary zone is considered as a pressure outlet, as shown in Figure 2. The SIMPLEC scheme, which is based on a predictor-corrector approach to enforce mass conservation, is used to coupling of pressure and velocity fields.

Green-Gauss cell-based method is used for Spatial Discretization. The second order discretization scheme is used for Pressure equations, second order upwind scheme are used for momentum, turbulent kinetic energy, specific dissipation rate. All equations' scaled residuals were aimed for 1e-05 convergence.

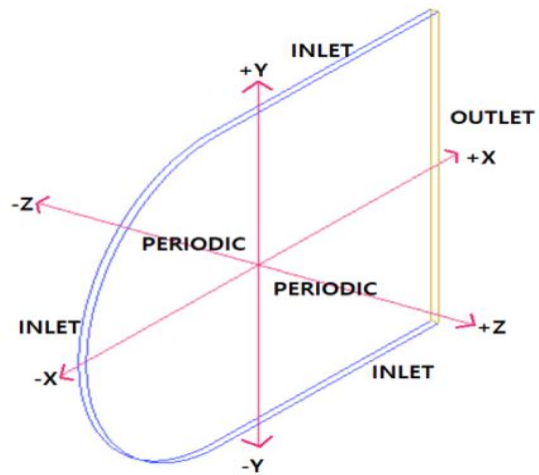

**FIGURE 2**: Boundary Conditions

### 2.2 Grid Independence Study

A sequence of coarse, medium and fine meshes have been created and analyzed at an AOA of 6.3°. Grid independence study has been performed to reduce the influence of the grid size on the computational results. We can see from Figure 3 that the $C_l$ values have settled at around 10 lakhs and are changing negligibly as we increase the size, so mesh with 10 lakh cells has been chosen for our study to reduce the computational time of the simulations.

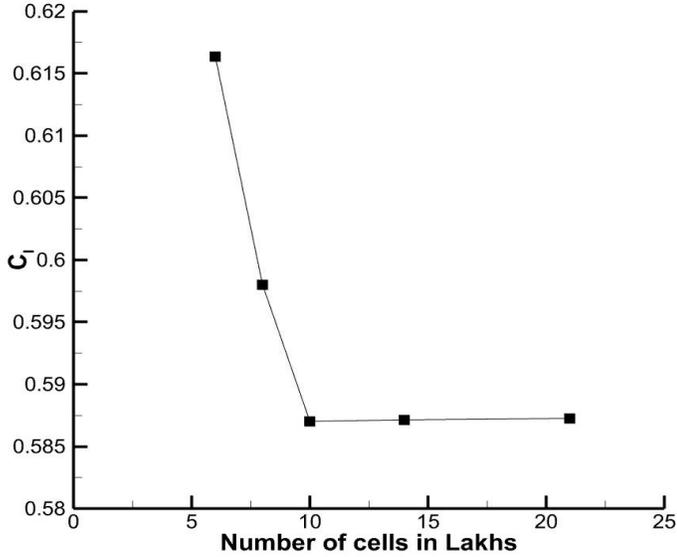

**FIGURE 3:** Grid Independence Study

### 2.3 Validation

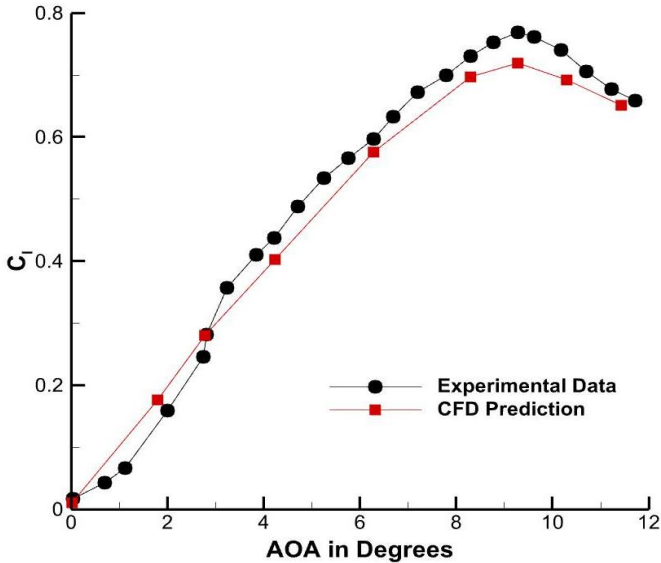

**FIGURE 4:** Lift Coefficient vs Angle of Attack

The validation of the CFD prediction is done with the experimental data of Winslow et al [24], and it is seen that the CFD prediction matches closely with the experimental data at all AOA with a maximum error of 6.49% at the $PL_\alpha$ i.e., 9.3°.

### 2.4 Definition of Protuberance

The undulations in the Leading Edge of the blade are known as Protuberance. In the present investigation sinusoidal protuberance is chosen as Swastika et al [25] concluded that sinusoidal protuberances are beneficial in the post stall region. Since the main motive in the present investigation is to improve the post stall performance, sinusoidal protuberance is chosen.

**TABLE 1:** Various configurations of Protuberances

| PAR | Amplitude, A |
|-----|--------------|
| 1   | 0.25c        |
| 3   | 0.083c       |
| 6   | 0.041c       |
| 9   | 0.027c       |
| 12  | 0.020c       |
| 18  | 0.013c       |
| 21  | 0.011c       |
| 27  | 0.0092c      |

Various configurations of the protuberances were investigated by keeping the pitch with a constant value of 0.25c and varying amplitude and this parameter is defined as Pitch-to-Amplitude ratio (PAR) and various PARs used for the numerical investigation are listed in Table 1.

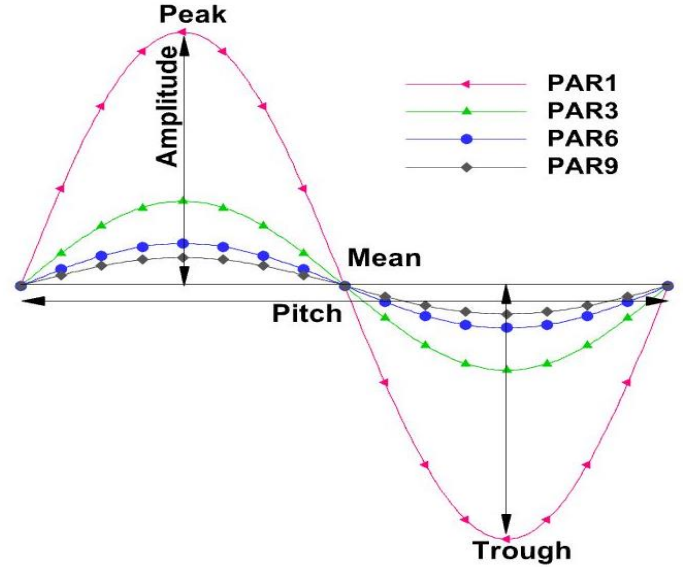

**FIGURE 5:** Definition of Protuberance

### 3. RESULTS AND DISCUSSION

A lift coefficient plot vs PAR plot for all the PAR cases and Baseline (Taking Baseline as PAR0) is shown in Figure 6. It can be observed that at the peak lift AOA of the baseline, the baseline case and all the higher PARs, i.e., PAR21, PAR24 and PAR27 show higher lift compared to all the other PAR cases. But after the baseline stalls, in the post-stall AOA of 11.4°, it can be observed that all the PAR cases show greater lift than the baseline case, with PAR3, PAR6 and PAR9 being the best cases. As the AOA is further increased to 13.6°, the baseline undergoes deep stall, and out of all the PAR cases, PAR3, PAR6 and PAR9 show the most promising performance. Hence to reduce the computational cost, for this study, it was decided to investigate only these three promising PAR cases. Further down the section, results of PAR3, PAR6 and PAR9 will be discussed.

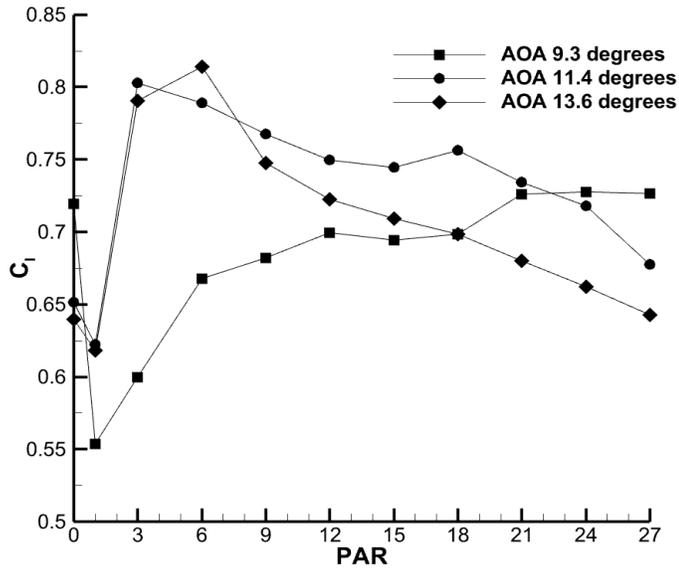

**FIGURE 6:** $C_l$ vs PAR cases at 9.3° (Stall AOA), 11.4° (Post-stall AOA), 13.6° (Deep-stall AOA).

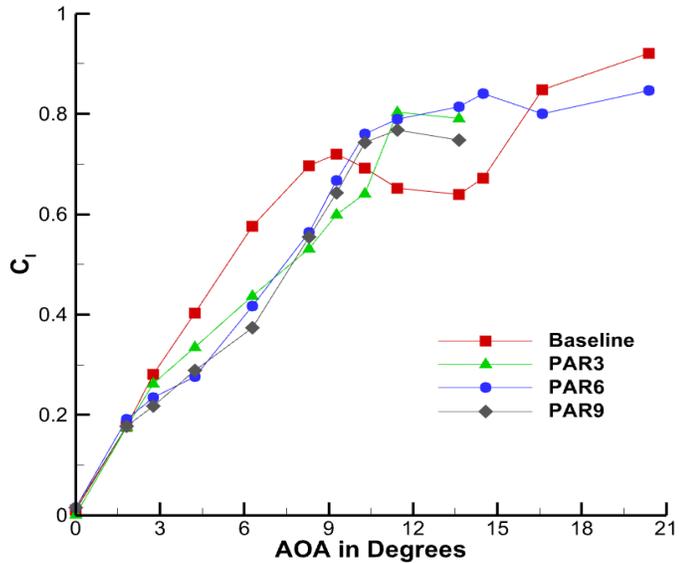

**FIGURE 7:** Comparison of $C_l$ vs AOA for baseline and effective PAR Cases.

For the baseline, $C_l$ rises linearly upto AOA 9° and stalls at AOA 9°. Further, $C_l$ drops upto AOA 13.6°. Whereas PAR3, PAR6 & PAR9 has greater $C_l$ compared to the baseline in the post-stall AOAs of the baseline. Also, it can be noted that all the modified cases have less $C_l$ in the pre-stall AOAs of the Baseline. The exact increase of $C_l$ in comparison with PAR cases for the post-stall AOAs is shown using Bar graph later.

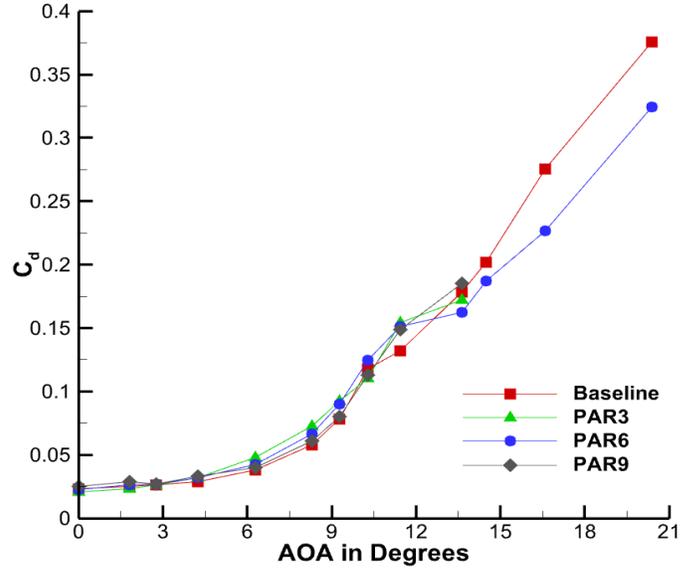

**FIGURE 8:** Comparison of Cd vs AOA for baseline and effective PAR Cases.

Figures 7 to 9 shows the comparison of the aerodynamic parameters for the baseline with PAR3, PAR6 & PAR9 cases. PAR3, PAR6 & PAR9 cases were found to be effective in the post stall region of the baseline. Figure 5 shows the comparison of $C_l$ vs AOA for the baseline and the PAR cases.

Similarly, Figure 8 shows the comparison of $C_d$ for the baseline and the PAR cases. $C_d$ is almost identical in the pre-stall AOAs and there is minor increase in Cd for the PAR cases in pre-stall AOAs. At 13.6° AOA, except PAR9, other two modified cases have less $C_d$ compared to the baseline. So, this also helps in the increase in the efficiency of the PAR cases. From Figure 9, it is observed that between AOA 0° and 2°, the baseline's and the protuberance cases' efficiency is identical, and between AOA 2° and 10°, a drastic drop in efficiency is observed, with the baseline having as much as 40% higher efficiency than the PAR cases, the reason for which will be explained later. But after the stalling angle of the baseline, the PAR cases have a higher efficiency as compared to baseline with PAR6 being the best performing case with as much as 39.6% more efficiency than the baseline at AOA 13.6°.

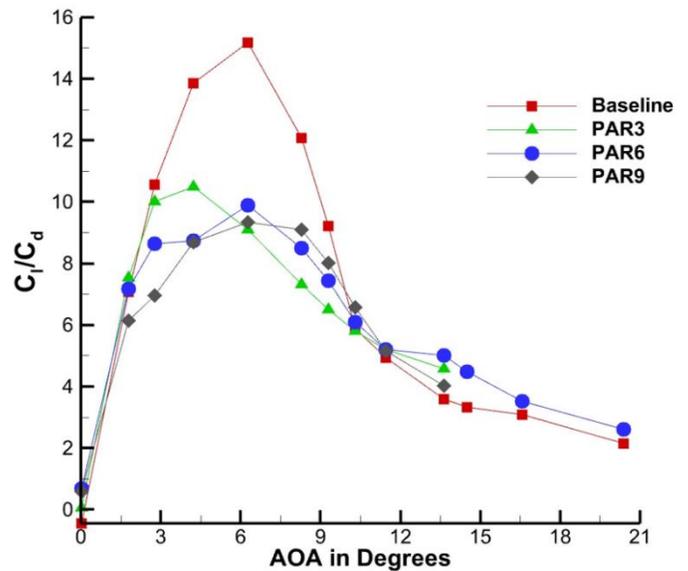

**FIGURE 9:** Comparison of $C_l/C_d$ vs AOA for baseline and effective PAR Cases

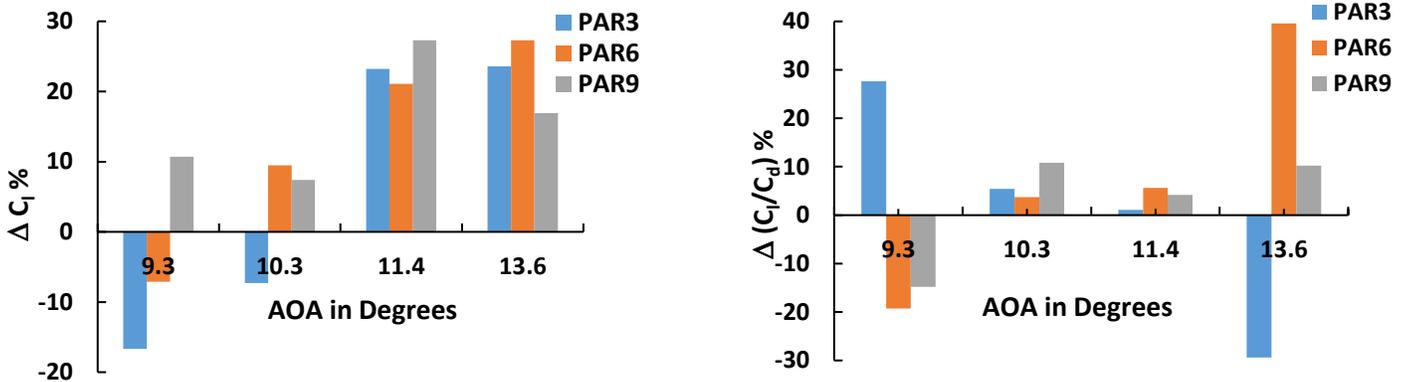

**Figure 10:** a) Percentage change in $C_l$ with respect to Baseline for PAR Case. b) Percentage change in $C_l/C_d$ with respect to Baseline for PAR Case

Results of the best performed modified cases in comparison to the baseline model in terms of $C_l$ and $C_l/C_d$ are shown above in Figures 10 (a) and 10 (b). As mentioned earlier, PAR of the protuberances in the leading edge of blade having the underlying profile of NACA 0009 airfoil is varied from PAR1 to PAR27. PAR3, PAR6 & PAR9 are found to have the most desirable characteristics in the post stall region compared to the other PAR cases. So, the variation of $C_l$, $C_l/C_d$ values are discussed in detail for PAR3, PAR6 and PAR9 cases without going in detail for the other PAR cases. All the above mentioned best performing modified blades have better $C_l$ in the post stall region compared to the baseline blade, but they have lesser lift at the peak lift angle of the baseline. At the peak lift angle (Stall AOA) of the baseline i.e., at 9.3°, lift coefficient of PAR3, PAR6 and PAR9 cases is reduced by 16.7%, 7.2% and 10.7% respectively. Also, efficiency is decreased in case of PAR6 and PAR9 by 19.3% and 14.8% respectively whereas in case of PAR3, it is increased by 27.6%. At 13.6°, PAR6 has highest lift with increase in lift around 27% while PAR3 and PAR9 cases have increase in lift around 23% and 18% respectively.

Comparing the effectiveness of PAR cases for efficiency at deep stall angle of attack of the baseline i.e., at 13.6°, PAR6 case has the best performance with increase in efficiency around 40%. While the other two PAR cases have less efficiency compared to PAR6 i.e PAR3 has 30% decrease in efficiency compared to the baseline whereas PAR9 has increase in efficiency of 10%. Overall, PAR6 has the best performance compared to the all-modified cases. PAR6 case and flow behavior associated with it is discussed in detail in further sections.

### 4. Result of Pressure Coefficient

#### 4.1. At Angle of Attack 9.3°

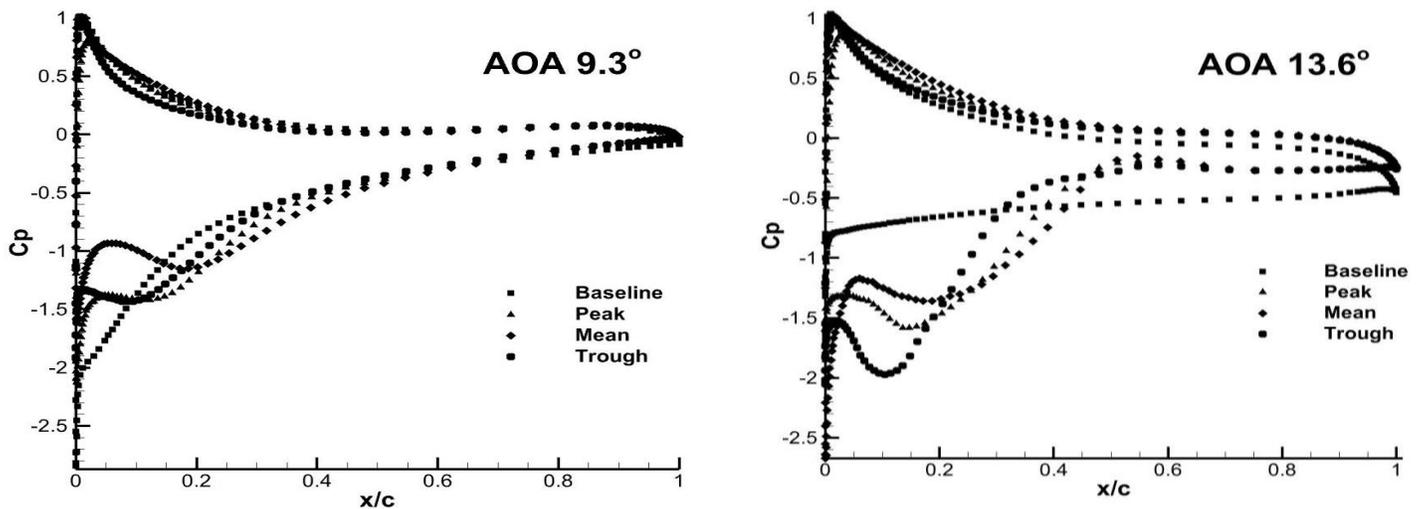

**Figure 11:** Comparison of Pressure distribution over Baseline and different sections of PAR6 case: a) 9.3° (left), b) 13.6° (right)

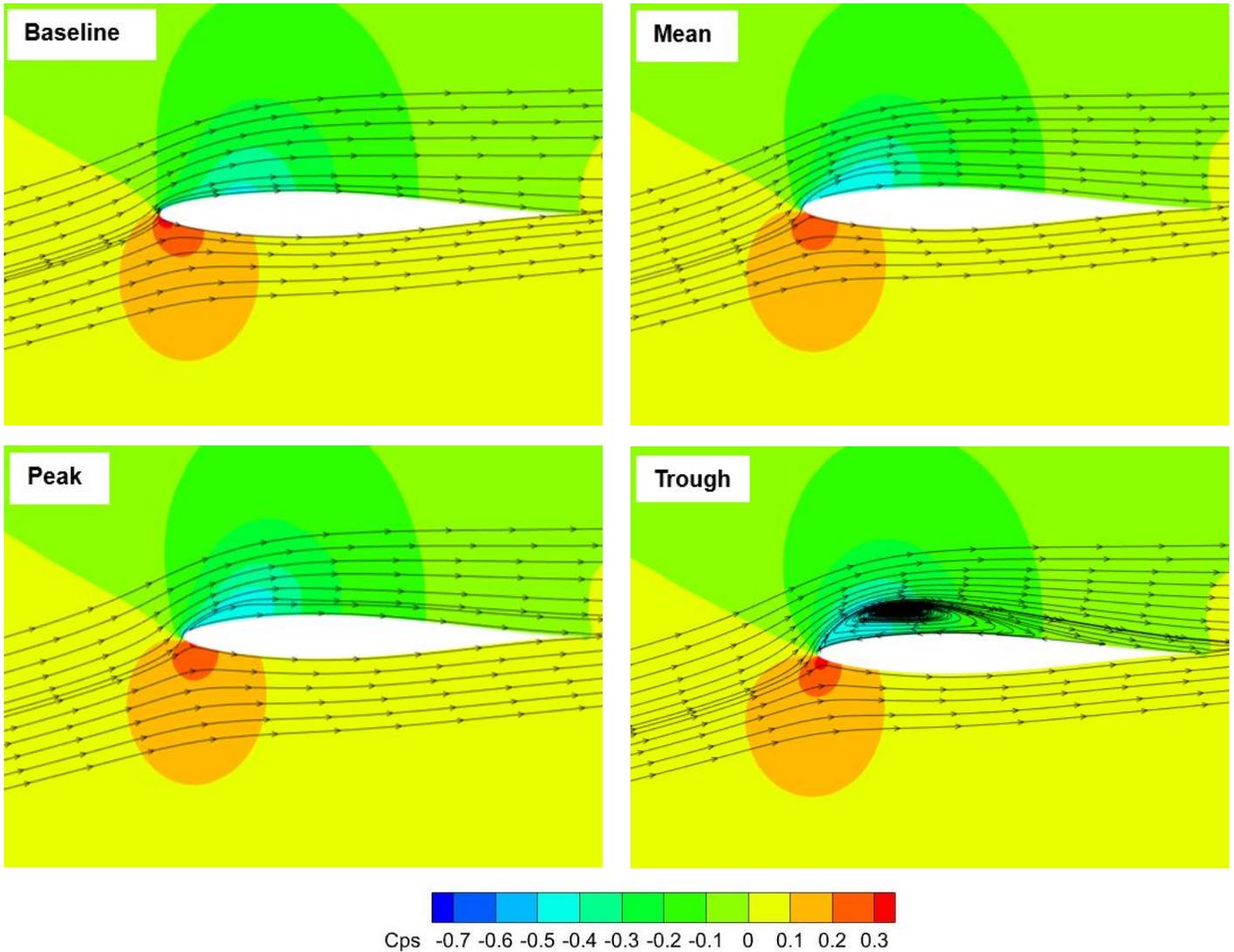

**Figure 12:** Flow visualisation through stream lines with Static Pressure Contours of (a) Baseline along with (b) Mean, (c) Peak, (d) Trough sections of PAR6 case at AOA 9.3°.

Plots of pressure distribution over the blade shows how load has distributed over its entire area. In the case of baseline, there is uniform distribution of load at every span wise location it differs in case of modified blades. Figure 11(a) shows the pressure distribution over baseline cross section in comparison with peak, mean and trough cross sections of PAR6 blade at the peak lift AOA of the baseline, i.e., 9.3°. It can be seen clearly that how wing loading changes at peak, mean and trough.

It is known that area under the curve of pressure distribution represents the lift. So, calculating the area under the curve for each cross section and doing the chord weighted average to get the overall area of the modified blade, it is observed to have 12.1% less compared to baseline. It is also evident as $C_l$ for PAR6 case is less compared to Baseline case at AOA 9.3°. It is important to investigate how the flow advances on the baseline as well as PAR6 case. This discussion will be further continued in the later sections.

### 4.2. At Angle of Attack 13.6°

Conducting the similar observations of pressure distribution over the blade at AOA 13.6°, it can be observed from Figure 11(b) that the pressure distribution of the baseline cross section has deteriorated compared to AOA 9.3°. As the AOA is increased to 13.6°, separation is found in the baseline which changes the pressure distribution which eventually leads to increase in pressure drag. Similarly, calculating the area under the curve as in the case of 9.3°, area under the curve for PAR6 case has 5% more area compared to the Baseline. Hence, more lift is obtained in case of PAR6.

### 5. Flow behavior over suction surface of blades

Figure 12 shows the flow visualization over the baseline and Peak, Mean, Trough section of PAR6 blades at peak lift angle of attack i.e., 9.3°. At the AOA 9.3°, flow is completely attached over the baseline blade which helps in producing the peak left. In Case of PAR6 blade, similar activity is observed over peak and mean section whereas in trough section there is a creation of recirculation region. This recirculation region creates losses and hence the PA R6 case has slightly lower lift compared to baseline blade. In Figure 12, flow visualization shown through streamlines is done on the static pressure contours. There are no much major differences in the pressure distribution over the baseline blade and PAR6 blade. As the AOA is increased to 13.6º, some interesting flow behaviors are observed.

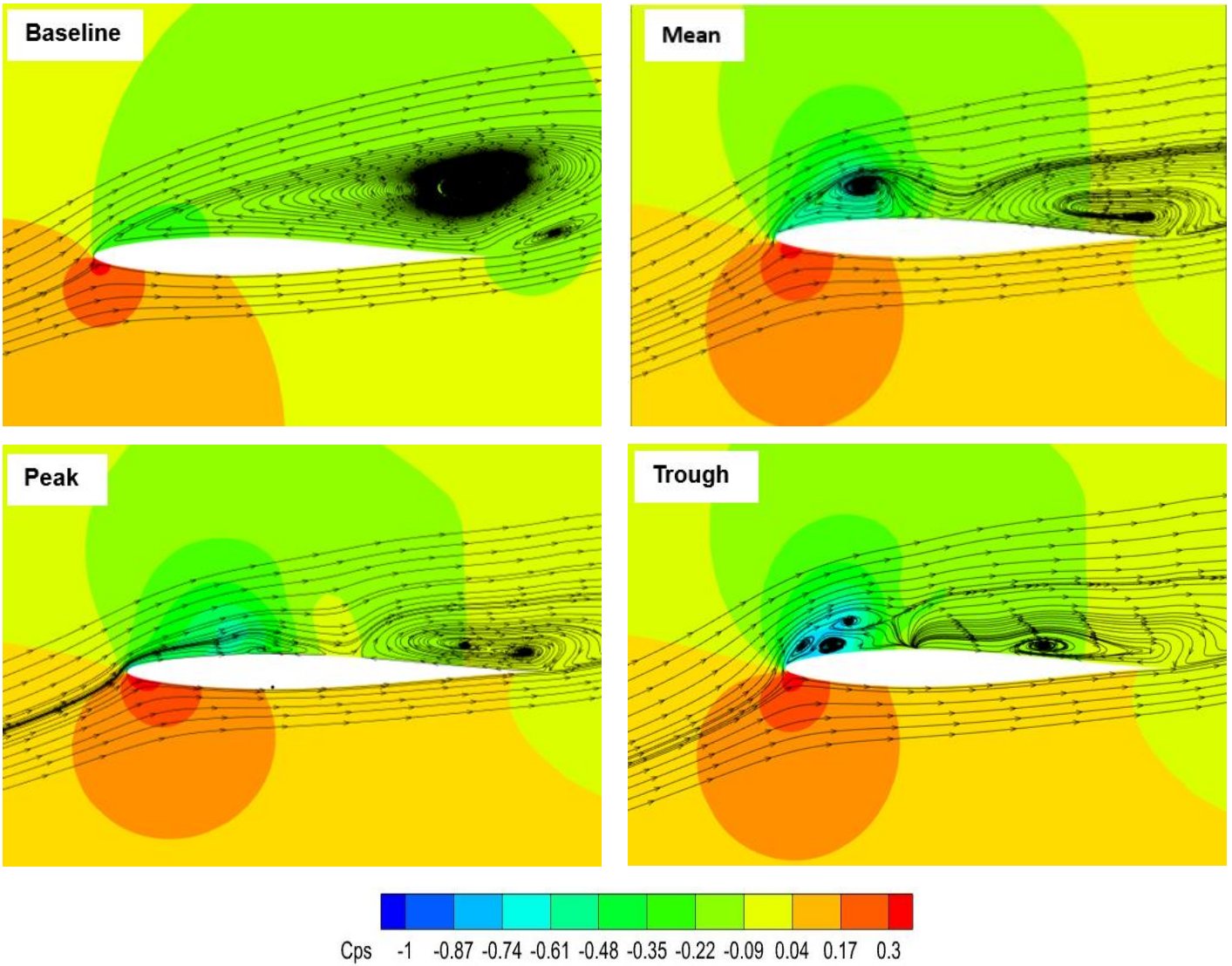

**Figure 13:** Flow visualisation through stream lines with Static Pressure Contours of (a) Baseline along with (b) Mean, (c) Peak, (d) Trough sections of PAR6 case at AOA 13.6º.

Figure 13 shows the flow visualization over the baseline and Peak, Mean, Trough section of PAR6 blades at deep stall angle of attack of baseline i.e 13,6º. Here, in case of baseline flow separation has taken place over the entire suction surface and massive separation region is formed which decreases the lift with drastic increase in pressure drag. This results in severe drop in aerodynamic efficiency. On the other hand, PAR6 case has different flow activity at Peak, Mean and Trough sections. As the streamlines are drawn with static pressure contours, this helps in pointing out the striking differences observed in the pressure distribution. On every section of PAR6, there is creation of low-pressure region at the leading edge. This is slightly more in case of trough region. So, it signifies that the flow is accelerated at the trough region.

In Figure 13(d), it can be seen how the accelerated flow has created the high strength recirculation regions which are opposite in directions to each other. These recirculation regions interact with each other to make sure the flow is attached to the surface. The flow is separated at the leading edge but it is quickly reattached due to these recirculating regions. Similarly, Peak and Mean section of PAR6 case has the recirculation regions. Recirculation regions are present throughout the end of the chord on the PAR6 case. The flow has become complex with the presence of numerous recirculation regions.

In overall, for PAR6, flow separation takes place near the leading edge but quickly reattaches downstream of the flow leading to the attachment of flow over the suction surface. This results in the preservation of lift and no drastic increase in drag is observed. Hence, retaining the aerodynamic efficiency. Since reattachment of the flow is observed on the modified blade, it is important to investigate the complex movement and interaction of recirculation regions in the flow to understand the mechanism behind the reattachment.

As seen in the case of PAR6 in Figure 13, there is complex flow behavior seen on the suction surface of the blade. So, in the further section a detailed spanwise flow study will be discussed to try to understand the complex flow behavior observed with the help of vorticity and total pressure loss contours.

## 6. Vorticity Contours

The flow on the suction surface of the blade in case of PAR6 has interesting, challenging flow features that can be difficult for the naive minds to understand. This complex flow behavior needs a detailed study with deep insights of flow physics. In this section an attempt is made to explain the mechanism behind the protuberances.

In case of PAR6, flow over the suction surface has combination of reattachment regions, recirculation regions with vortices which changes distinctly as the flow progresses downstream. The strength of vortices also changes massively from the leading-edge of the blade to the trailing edge of the blade. In this section, x-vorticity contours have been taken on chordwise slices at different chord locations. Many chordwise slices of x-vorticity are studied and the slices with more activity are shown and discussed further. Figure 14 shows the vorticity contours of the PAR6 blade at chordwise locations at 0.04m, 0.08m and 0.3m. In Figure 14 (a), the chordwise slice is made at 0.04m. At this location, the flow has not swept the entire leading edge due to the modification of the leading edge. The peak region has flow over it while in the trough region, the flow is just about to reach.

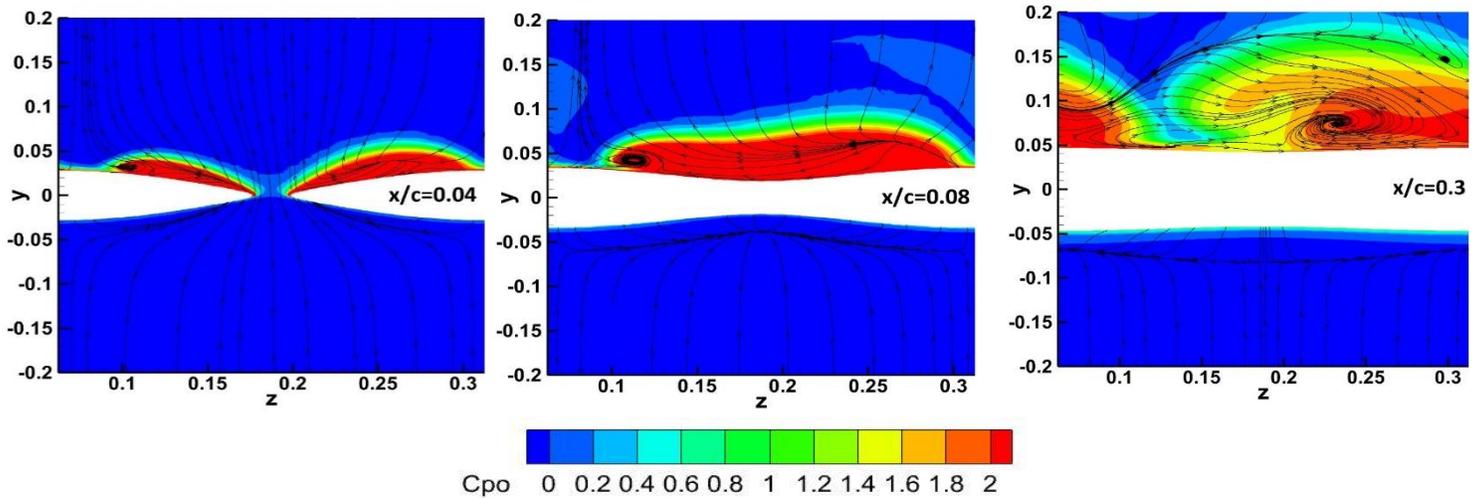

**Figure 14:** a) Chordwise slice of PAR6 blade at 0.04m (left), b) Chordwise slice at 0.08m (centre), c) Chordwise slice at 0.3m (right) at AOA 13.6º.

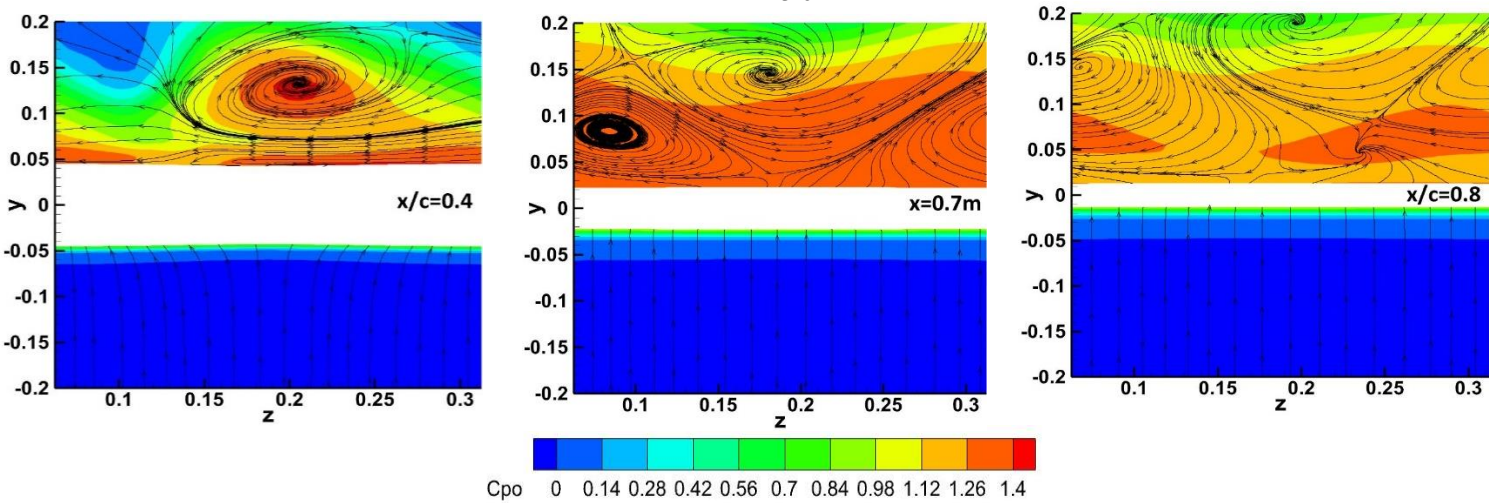

**Figure 15:** a) Chordwise slice of PAR6 blade at 0.4m (left), b) Chordwise slice at 0.7m (centre), c) Chordwise slice at 0.8m (right) at AOA 13.6º.

So, the flow is in more ease at the trough region. Further the downstream of the flow, the flow from the trough region rushes towards the peak region of the flow. This movement of the flow creates distinct flow behaviors which helps in attaching the flow to the blade and improving the aerodynamic performance. By thoroughly investigating Figure 14(a), the presence of two vortices with high strength can be seen. These two vortices have opposite signs signifying that those vortices are counter rotating vortices. Also, there is a beginning of the formation of recirculation region. As the flow advances, this recirculation region grows which is seen in the Figure 14(b). Figure 14(b) shows the x-vorticity contour at chordwise location of 0.08m.

Here, the flow has swept in the entire span wise. Compared to the vortices in Figure 14(a), vortices in Figure 14(b) have less strength. The recirculation region in the left side has grown bigger. Also, there is a hint of formation of the bigger recirculation in the right side. So, further taking the chordwise slice at 0.3m, it can be observed that the strength of the vortices is further reduced. Interestingly, as hinted earlier, there is formation of massive recirculation region. These vorticity contours unfurl the reason behind the complex flow behavior happening over the suction surface of the blade due to leading edge modification. Here, we have mainly two activities which need more attention. Firstly, the formation of the vortices at the leading edge which dies steadily further downstream and secondly, the formation of the recirculation regions. These recirculation regions change their size and location as the flow advances and results in the formation of new recirculation regions. Formation of new recirculation regions do not have any pattern which indicates the irregularity of the flow field formed due to the modification of the leading edge. Since, vortices go on weakening their strength, it seems there is no point in investigating further downstream of 0.3m position. But the presence of massive recirculation region in Figure 14(c) along with another smaller recirculation region needs further investigation. Investigating the vorticity contours at 0.4m, 0.7m, 0.8m also shows the presence of recirculation regions. The recirculation region formed at 0.3m goes on growing further downstream and is preserved till the trailing edge of the blade. This is evident from Figures 15(a), 15(b), 15(c).

As mentioned earlier in the discussion, the strength of vortices in Figure 15 is too less compared to the vortices that are formed in Figure 14. In Figure 15(b), there is presence of Laminar Separation Bubble, which is normally observed in Reynolds number ranging from 20000 to 50000. In overall, observing Figures 14 and 15, it can be concluded that the flow over the suction surface of the blade is highly dynamically unstable, irregular, and complex.

The presence of numerous recirculation regions over the suction surface indicates the rearward movement of the flow once it reaches the trailing edge of the blade. In case of the baseline blade, the flow leaves the trailing edge of the blade casually. Whereas in case of the modified blade (PAR6), some of the flow comes back and results in the formation of these recirculation regions.

## 7. Boundary Layer Thickness (BLT)

Boundary layer thickness is a significant parameter to see the flow behavior over the airfoil. Boundary layer is too thin at low AOAs which leads to smooth flow around the airfoil. As the AOA is increased beyond the stall angle, thickness of boundary layer goes on increasing which results in the flow separation with high pressure drag. This results in the abrupt stall characteristics of the airfoil. Boundary layer thickness is calculated up to the region where 99.99% of free stream velocity is achieved. At every 10% of chord, chord wise sections are created and mass weighted average velocity of that section is calculated. The height of the sections are 2 times the length of the chord above and below. The boundary layer thickness is then calculated where the mass weighted average velocity reaches 99%.

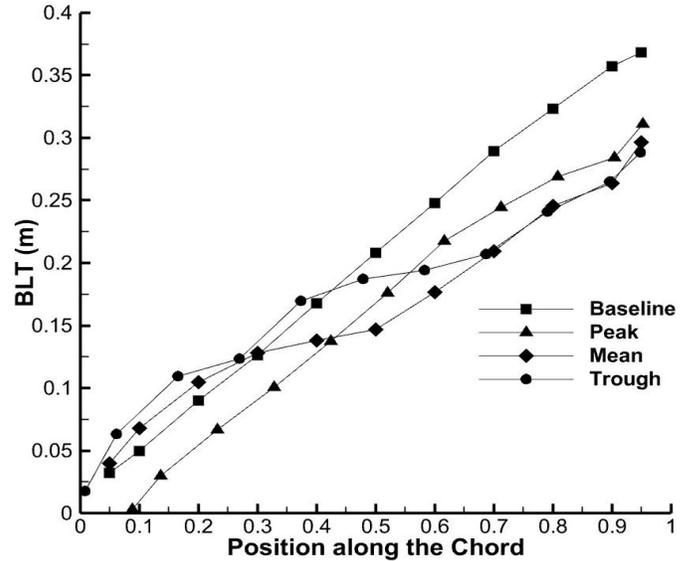

**Figure 16:** Boundary layer thickness variation along the chord for baseline and different sections of PAR6 at AOA 13.6°

Figure 16 represents how the boundary layer thickness varies along different positions of the chord of baseline and different sections of PAR6. It can be observed that for the baseline the thickness of the boundary layer has increased linearly along the chord. In case of the PAR6 sections, thickness of boundary layer has not increased linearly and has shown different activity. As already seen in Figure 13, there is presence of recirculation regions between 0 to 0.2m for the mean and trough sections. Hence, in Figure 16, between 0 and 0.2m the BLT is more for these two sections compared to the baseline, whereas in the peak region the flow is still attached having thin BLT as observed in Figure 16. Also, from Figure 11(b), it can be seen that all the sections of PAR6 have better wing loading compared to the baseline. Between 0.4m to 1m, the BLT of the baseline is more which is an indication of flow separation which is evident from Figure 13(a). Overall, BLT of the PAR6 is less compared to the baseline. So, it has lift and less drag at this deep stall AOA of the baseline.

## CONCLUSIONS

- The baseline blade stalls at around 9.3° leading to deteriorating lift, drastic increase in drag, thereby leading to drastic drop in efficiency.
- The Pitch to Amplitude ratio of the Protuberance is varied from PAR1 to PAR27. Among the cases investigated, PAR3, PAR6 & PAR9 are the better performing cases compared to other cases.
- PAR1 case is the worst case in terms of $C_l$, $C_d$. Since the geometry of the PAR1 has steep curves, flow gets into condition where turbulent energy of the flow is undesirable.
- PAR6 case is found to be the best performing case in the range of PARs investigated, showing better post-stall performance than the baseline and the other PAR cases. It has 39.6% and 27.3% more lift and efficiency respectively compared to the baseline at its deep stalling AOA.
- The mechanism behind the better performance of PAR6 in the post-stall region is due to the formation of counter-rotating vortices that helps to mix the low momentum flow with the high momentum flow. This helps in delaying in flow separation over the suction surface.
- In UAVs, military aircrafts etc., which operates at high angle of attacks, use of protuberance maybe beneficial as it helps to delay flow separation, and helps to maintain high lift even at high AOAs.